\documentclass[conference]{IEEEtran}
\IEEEoverridecommandlockouts

\usepackage{amsmath}
\usepackage{latexsym}
\usepackage{amssymb}
\usepackage{epsfig}
\usepackage{moreverb}
\usepackage{rotating}
\usepackage{enumerate}
\usepackage{graphics, graphicx,wrapfig}
\usepackage{fancybox}
\usepackage{float}
\usepackage{url}
\usepackage{tabularx}
\usepackage{algorithmic}
\usepackage{algorithm}
\usepackage{array}
\usepackage{booktabs}
\usepackage{multirow}

\usepackage{listings}
\usepackage{hyperref}
\usepackage{array}
\usepackage{balance}
\usepackage{subcaption}
\def\BibTeX{{\rm B\kern-.05em{\sc i\kern-.025em b}\kern-.08em
    T\kern-.1667em\lower.7ex\hbox{E}\kern-.125emX}}
\begin{document}

\title{Leveraging MTD to Mitigate Poisoning Attacks in Decentralized FL with Non-IID Data}

\author{
    \IEEEauthorblockN{Chao Feng\IEEEauthorrefmark{1}, Alberto Huertas Celdrán\IEEEauthorrefmark{1}, Zien Zeng\IEEEauthorrefmark{1}, Zi Ye\IEEEauthorrefmark{1}, Jan von der Assen\IEEEauthorrefmark{1}, \\ G\'er\^ome Bovet\IEEEauthorrefmark{2}, Burkhard Stiller\IEEEauthorrefmark{1}}
    \IEEEauthorblockA{\IEEEauthorrefmark{1}Communication Systems Group, Department of Informatics, University of Zurich, 8050 Zürich, Switzerland \\{[cfeng, huertas, vonderassen, stiller]}@ifi.uzh.ch, [zien.zeng, zi.ye]@uzh.ch}
    \IEEEauthorblockA{\IEEEauthorrefmark{2}Cyber-Defence Campus, armasuisse Science \& Technology, 3602 Thun, Switzerland gerome.bovet@armasuisse.ch}
}

\DeclareRobustCommand*{\IEEEauthorrefmark}[1]{%
  \raisebox{0pt}[0pt][0pt]{\textsuperscript{\footnotesize #1}}%
}

\maketitle

\begin{abstract}

Decentralized Federated Learning (DFL), a paradigm for managing big data in a privacy-preserving and distributed manner, is vulnerable to poisoning attacks where malicious clients tamper with data or models. Current defense methods often assume Independently and Identically Distributed (IID) data across participants, which is unrealistic in real-world applications. In more realistic non-IID contexts, existing defensive strategies face challenges when distinguishing between models that have been compromised and those that have been trained on heterogeneous data distributions (non-IID), leading to diminished efficacy. In response, this paper proposes a framework that employs the Moving Target Defense (MTD) approach to bolster the robustness of DFL models. By continuously modifying the attack surface of the DFL system, the framework aims to mitigate poisoning attacks effectively. The proposed solution includes both proactive and reactive modes, utilizing a reputation system that combines metrics of model similarity and loss, alongside various defensive techniques. Comprehensive experimental evaluations indicate that the MTD-based mechanism significantly mitigates a range of poisoning attack types across multiple datasets with different federation topologies.
\end{abstract}


\section{Introduction}
\label{sec:intro}
The rapid expansion of the Internet-of-Things (IoT) has generated a staggering 402.74 million Terabytes of data every day~\cite{explodingtopics2024}. Federated Learning (FL), characterized by its distributed and privacy-preserving attributes, has emerged as a powerful paradigm for addressing the challenges associated with big data~\cite{mcmahan2017communication}. Nevertheless, the vanilla FL paradigm relies on a central server for global model aggregation, which introduces a processing bottleneck and the risk of single-point failures within the system. In response, the Decentralized FL (DFL) paradigm has garnered increasing attention, as it eliminates the client-server boundary, thereby significantly enhancing the system's stability and robustness~\cite{beltran2023decentralized}.

The decentralized nature of DFL renders it susceptible to various forms of adversarial attacks, especially poisoning attacks~\cite{feng2024dart}. Without a central server that regulates, authenticates, and authorizes the federation, malicious participants can exploit this structure by manipulating local data or models. This manipulation enables them to disseminate malicious models across the federation, potentially leading to the introduction of targeted malice or a deterioration in the integrity and robustness of the DFL model.

Most of the conventional defenses designed to mitigate poisoning attacks in FL, such as Krum~\cite{blanchard2017machine}, depend on the similarity or distance metrics between models. These approaches implicitly presume that benign models exhibit a high degree of similarity, predicated on the assumption that their training datasets are independent and identically distributed (IID). However, this assumption is often unrealistic, as real-world scenarios frequently present significant variations in data distribution across nodes. The challenge posed by non-IID data results in increased distances between models and diminished similarity, thereby undermining the effectiveness of conventional defense mechanisms in non-IID contexts~\cite{singh2023fair}.

Movement Target Defense (MTD) is an emerging cybersecurity paradigm for mitigating various attacks~\cite{zheng2019survey}. By dynamically altering the attack surface, either actively or passively, MTD increases the complexity and cost of successfully executing a cyberattack~\cite{cai2016moving}. A limited number of studies have endeavored to employ MTD to mitigate security concerns within DFL~\cite{martinez2024mitigating, feng2023voyager}. However, their performance is diminished in non-IID environments. Thus, there is a need to apply MTD to enhance the security and robustness of DFL models, particularly in defending against poisoning attacks.

This work proposes an MTD-based approach to dynamically modify the attack surface of DFL systems and improve their resilience and security. Therefore, the main contributions of this paper are: (1) Designing an MTD-based approach to mitigate poisoning attacks in DFL. This approach utilizes a dual-reputation score system based on model similarity and loss, adopts proactive and reactive modes, and incorporates dynamic topology and aggregation strategies. (2) Employing the DBSCAN clustering algorithm to adjust the triggering threshold for MTD dynamically. (3) Implementing a prototype of the proposed MTD method on a real DFL platform~\cite{beltran2024fedstellar}. (4) Conducting extensive experiments to compare the proposed algorithm with state-of-the-art methods, including scenarios with fully connected, ring, and star topologies, as well as untargeted poisoning and backdoor attacks on MNIST, FashionMNIST, and CIFAR10 datasets. The results demonstrate that the proposed MTD-based algorithm not only effectively mitigates poisoning attacks, outperforming existing methods, but also improves model performance on non-IID data.

The remainder of this paper is organized as follows: Section~\ref{sec:relatedwork} provides a summary and review of research utilizing MTD to address AI security challenges. Section~\ref{sec:problem} outlines the problems this paper aims to solve. Section~\ref{sec:approach} details the design of our proposed MTD-based approach. Subsequently, Section~\ref{sec:exp}  evaluates and compares the effectiveness of our method with state-of-the-art approaches in mitigating poisoning attacks. Section~\ref{sec:discussion} discusses the experimental results, and Section~\ref{sec:conclusion} concludes the paper by summarizing key contributions and highlighting future research opportunities.

\section{Background and Related Work}

\label{sec:relatedwork}

\begin{table}[t]
\centering
    \caption{MTD Applications in AI Security}
    \label{tab:related_mtd}
    
    \setlength{\tabcolsep}{3pt}
    \resizebox{\columnwidth}{!}{%
    \begin{tabular}{@{} llllll @{}} 
    \hline \textbf{Reference} &  \textbf{Scenario} & \textbf{Method} & \textbf{Target} & \textbf{Timing} & \textbf{Threat}\\
    \hline 
    \cite{roy2019moving} 2019 & ML & Diversity & Model Ensemble & Proactive & Adversarial Attack \\
    \cite{amich2021morphence} 2021 & ML & Diversity & Model Pool  & Reactive & Adversarial Attack \\
    \cite{qiu2021mt} 2021 & ML  & Shuffling & Model Selection & Proactive & Backdoor \\
    \cite{sengupta2019mtdeep} 2019 & ML  & Diversity  & Model Ensemble & Proactive & Adversarial Attack \\
    \cite{zhou2021augmented} 2021 & CFL & Shuffling & Client Selection & Proactive & Poisoning Attack \\
    \cite{feng2023voyager} 2023 & DFL & Shuffling & Topology & Reactive & Poisoning Attack \\
    \textbf{This work} &   DFL & Shuffling & Topology +  & Proactive \& & Poisoning Attack \\  
    & & & Aggregation &  Reactive  
\\\hline
    \end{tabular}
    }
\end{table}


Robust aggregation functions are the most commonly adopted defense approaches in FL against poisoning attacks~\cite{feng2024dart}. Krum~\cite{blanchard2017machine} identifies the update that is nearest to the average of the remaining updates, deliberately excluding outliers from consideration. Median~\cite{yin2018byzantine} employs the median value of the updates, which exhibits reduced sensitivity to outlier effects. TrimmedMean~\cite{yin2018byzantine} calculates the updates' mean after excluding a predetermined proportion of extreme values. Nonetheless, these defenses predominantly utilize static strategies typically formulated for the conventional FL paradigm rather than for DFL. These static defenses are ineffective in addressing poisoning attacks targeting DFL systems, especially in non-IID environments~\cite{feng2024dart}.

\tablename~\ref{tab:related_mtd} provides an overview of the existing literature on applying MTD strategies in the context of  Artificial Intelligence (AI) Security to address diverse cyber threats. In the traditional AI field, MTD has been employed to mitigate adversarial samples and poisoning attacks. \cite{roy2019moving} proposes an MTD approach to enhance Deep Neural Network (DNN) robustness by dynamically changing model parameters during inference. This makes it harder for attackers to craft compelling adversarial examples, improving resilience against white and black-box attacks. Morphence \cite{amich2021morphence} is an MTD strategy generating diverse student models from a base model with random weight perturbations. Randomly selecting a student model for each inference reduces the adversarial attack success rate while maintaining accuracy. MT-MTD \cite{qiu2021mt} presents an MTD framework protecting DNNs against backdoor attacks in edge computing. It divides training data, has an attacker train each dimension, and the defender randomly selects dimensions for consensus, increasing attack cost and maintaining accuracy. MTDeep \cite{sengupta2019mtdeep} enhances neural network robustness using MTD by dynamically changing model architecture, making it harder for attackers to craft adversarial attacks.

However, there is limited research on utilizing MTD methodology to address the security challenges of FL systems. \cite{zhou2021augmented} enhances FL security with MTD by introducing an augmented dual-shuffle mechanism. This dynamically changes the attack surface by shuffling client selection and model aggregation, improving system security against various poisoning attacks. Voyager \cite{feng2023voyager} employs a reactive MTD-based aggregation protocol to bolster the resilience of DFL against poisoning attacks. This is achieved by dynamically manipulating the connectivity of the network topology. However, its efficacy remains to be validated in non-IID scenarios. 

In summary, MTD has been empirically demonstrated to address a wide range of security vulnerabilities, including poisoning attacks within the realm of AI security. Nonetheless, there remains a significant gap in the research regarding the application of MTD to counteract poisoning attacks in FL, particularly in DFL. This gap underscores the urgency of the topic and the need for further research. Therefore, this paper is motivated by the need to explore methodologies for mitigating various forms of poisoning attacks in non-IID environments through a dynamic approach.

\section{Scenario and Adversarial Setup}
\label{sec:problem}
This section analyzes the challenges of mitigating poisoning attacks, particularly highlighting the distinct behaviors these attacks exhibit in IID and non-IID data environments. Additionally, it examines the adversarial setup by considering the attacker's objectives, knowledge, and capabilities. This analysis provides a foundation for understanding the specific problem and scenario addressed in this paper.

\subsection{Poisoning Attacks in non-IID Environments}
Traditional aggregation functions utilized in FL systems, such as FedAvg and Krum, are based on the assumption of model homogeneity. This homogeneity is derived from two factors. First, the architectural configuration of the models must be uniform, meaning that the dimensions of the parameter tensors are identical. This uniformity enables the computation of statistical measures such as the mean or median of the model parameters or gradients. Second, it is assumed that the models are trained on homogeneous data samples, specifically that the local datasets at each client are IID. As a result, the local models across clients exhibit a significant similarity, facilitating the exclusion of potentially malicious models based on model similarity or distance metrics. However, this second assumption becomes invalid in scenarios characterized by non-IID data distributions. The lack of IID conditions leads to diminished model similarities, complicating identifying and excluding malicious models based on similarity assessments.


To examine the impact of non-IID data on the similarity of local models within the DFL, this work trained three local models and calculated their pairwise similarities utilizing the MNIST~\cite{mnistwebsite} dataset under both IID and non-IID conditions. The similarity measurements were derived using the cosine similarity metric, as delineated in Equation~\ref{eq:cossim}. 

\begin{equation}
\text{Cosine Similarity} = \frac{\mathbf{A} \cdot \mathbf{B}}{\|\mathbf{A}\| \|\mathbf{B}\|}
\label{eq:cossim}
\end{equation}

where $A$ and $B$ are the model parameter vectors.

\begin{figure}[t]
    \centering
    \subfloat[\centering IID Data]{{\includegraphics[height=4.2cm]{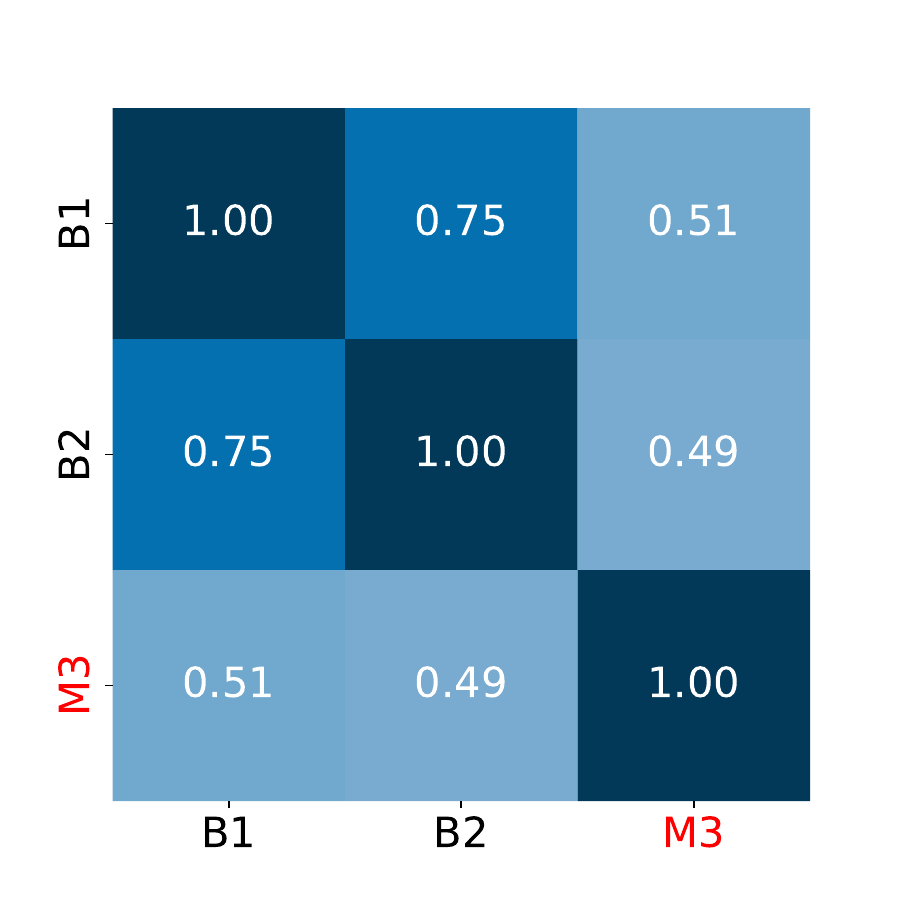} }}%
    \subfloat[\centering non-IID Data]{{\includegraphics[height=4.2cm]{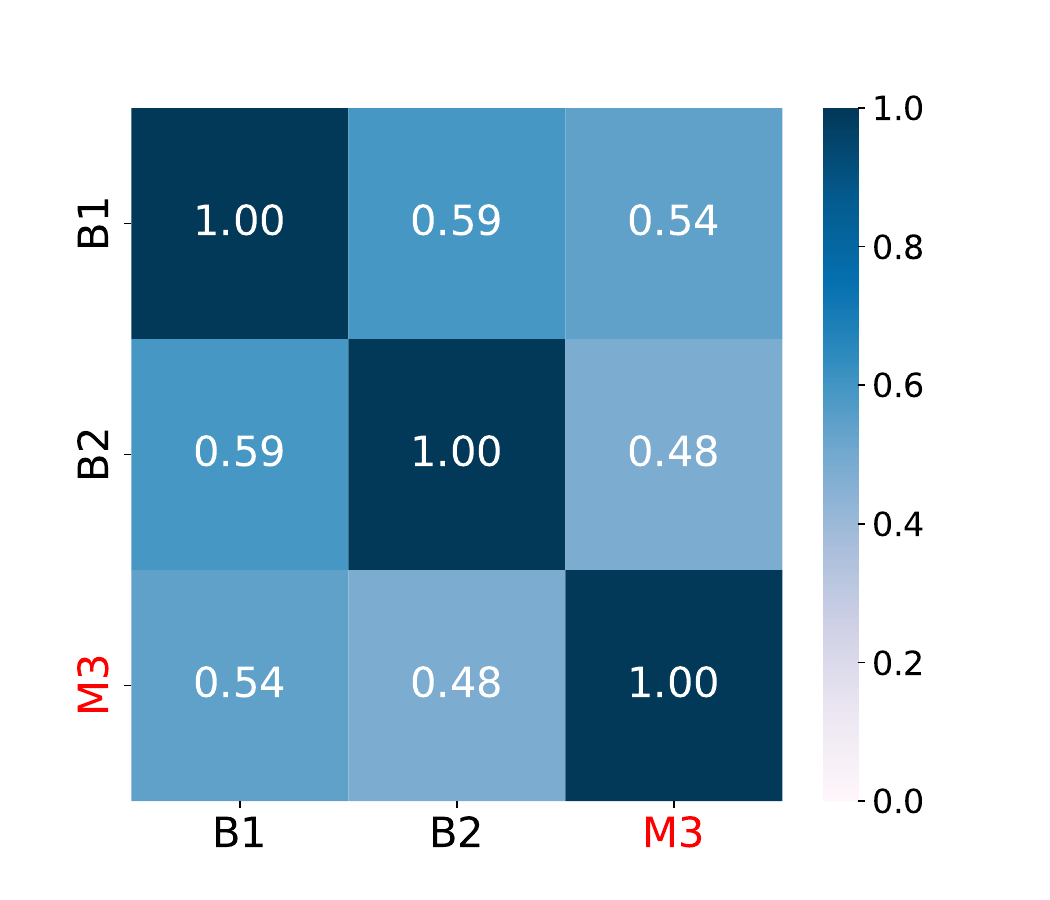}}} \\%
   {\footnotesize [Malicious Model ({M\textit{i}}), Benign Model (B\textit{j})]}         
   \caption{Pairwise Model Similarly in MNIST Dataset with Different Data Distribution}
    \label{fig:modelsim}    
\end{figure}

As illustrated in \figurename~\ref{fig:modelsim}, in both experimental configurations, nodes 1 and 2 were trained on normal data, whereas node 3 was malicious, having undergone a Label Flipping attack on its local dataset. In \figurename~\ref{fig:modelsim} (a), the similarity score between the benign nodes (B1 and B2) was determined to be 0.75, whereas the similarity between the benign and malicious nodes was approximately 0.5, indicating a significant disparity that suggests the potential effectiveness of similarity- or distance-based algorithms. Conversely, when employing a non-IID data distribution, as depicted in \figurename~\ref{fig:modelsim} (b), the similarity between the benign nodes decreased to 0.59, while the similarity between benign and malicious nodes increased to 0.54. This shift reduces the algorithm's ability to differentiate malicious nodes based on similarity.

\subsection{Adversarial Setup}
\label{sec:threatmodel}
This section examines poisoning attacks on DFL through the lens of the attacker's objectives, knowledge, and capabilities.

\textbf{Adversary Objective.} 
For poisoning attacks on DFL systems, the primary objective of the adversary is to undermine the robustness of the DFL model, thereby diminishing the overall efficacy of the aggregated model or introducing a specific backdoor. The potential impact of these actions is significant. Specifically, the attacker seeks to mislead the training process of the aggregated model by manipulating either the data or the model during the training phase. This manipulation ultimately steers the aggregated model toward a predetermined outcome. In the context of untargeted attacks, the attacker aims for the aggregated model to fail in generating reliable predictions. Conversely, in targeted attacks, the intention is to embed a specific backdoor within the aggregated model.

\textbf{Adversary Knowledge.} 
The attackers' knowledge is confined to their local environment, and they lack knowledge beyond this scope. They can leverage local training data and models, and they may be aware of the models shared by neighboring nodes. However, they do not have access to the local data or reputation scores associated with benign nodes.

\textbf{Adversary Capabilities.} 
Since the attackers have local knowledge of the training data and models, they can tamper with either the data (data poisoning) or the models (model poisoning), depending on the attack strategy employed by the attackers. However, their influence is constrained to the local environment, which means they can only indirectly impact the models of benign nodes by disseminating malicious model updates rather than directly altering the models of these benign nodes. An adversary can manipulate several malicious nodes, measured by the Poisoned Node ratio (PNR). This allows the attacker to execute attacks from multiple nodes simultaneously, thereby enhancing the likelihood that the benign models will be compromised.

\begin{figure}[t]
    \centering
    \includegraphics[width=\linewidth]
    {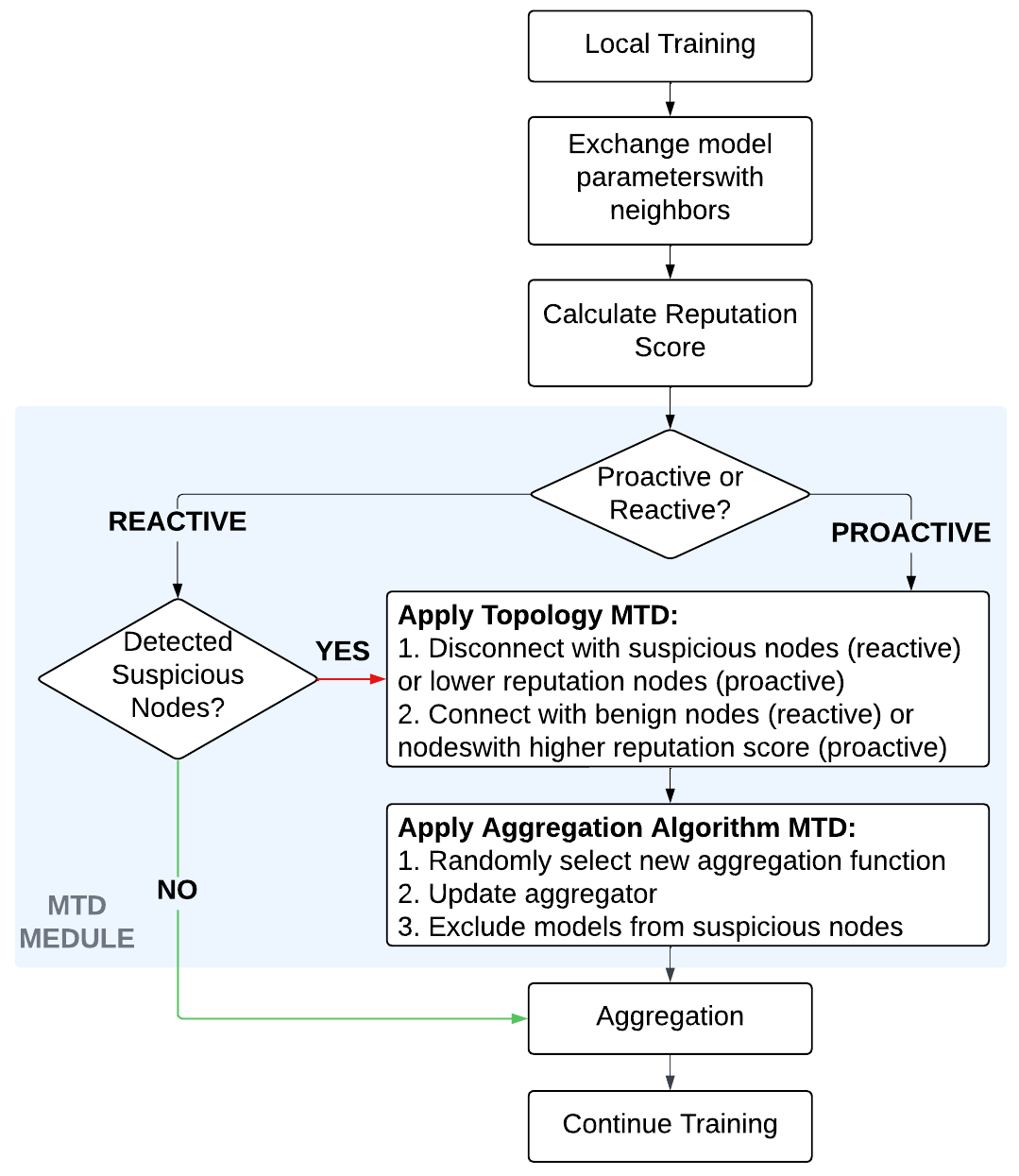}
    \caption{Overview of the MTD-based Approach}
    \label{fig:workflow}
\end{figure}

\section{MTD-based Approach}
\label{sec:approach}
This section describes the MTD-based approach used to defend against poisoning attacks in DFL, whose overall workflow is shown in \figurename \ref{fig:workflow}. Since DFL has no central controller to manage the security of the entire federation, all decisions are made locally at the nodes. 

Overall, each round of model training in DFL consists of five steps: 
\begin{itemize}
    \item Local Training: Each node trains a local model on its respective local dataset.
    \item Exchange Model Parameters: Each node sends its local model parameters to neighboring nodes and receives model parameters from them.
    \item Calculate Reputation Score: Each node evaluates the reputation score of the received models based on model similarity and loss on its local validation dataset. This evaluation helps in detecting suspicious nodes.
    \item Apply MTD Strategies: Proactively or Reactively apply dynamic topology and aggregation MTD strategies to mitigate poisoning attacks. 
    \item Model aggregation: Nodes aggregate models received from neighbors and their local model to update and improve their own model.
\end{itemize}
Each iteration of the training round is conducted in a cyclical manner until the model reaches convergence. This process ensures that the defense system can dynamically adapt to potential threats, maintaining the integrity and performance of the DFL models.

\subsection{Reputation System}
The reputation score serves a critical function, acting both as a trigger for the MTD module and as a foundational element for dynamic strategies. As shown in \figurename \ref{fig:modelsim}, a reputation system based solely on model similarity is insufficient, especially in non-IID environments. Therefore, this study proposes a dual reputation scoring system designed to leverage the strengths of DFL nodes, which can effectively utilize both local models and local datasets. This reputation system employs model similarity and validation set loss metrics to compute the reputation score for models shared from neighboring nodes. This reputation score is utilized locally to assess neighboring nodes' trustworthiness and determine whether the current node is under attack.

\textbf{Model Similarity.}
Cosine similarity is a commonly employed metric for quantifying the similarity between local and received neighbor models~\cite{feng2023voyager}. Cosine similarity quantifies the cosine of the angle between two non-zero vectors situated within an inner product space. As defined in Equation \ref{eq:cossim}, the cosine similarity values are constrained within the range of [-1, +1].

\textbf{Neighbor Model Loss.}
In addition to model similarity, the loss of the received model on the local validation dataset is also evaluated. This loss measurement helps determine models that do not perform well on the local data, further aiding in identifying malicious nodes. 

The model similarity and loss of neighboring models are integrated into a two-dimensional vector, forming the reputation system's metrics. The reputation score serves as the basis for judging the trustworthiness of neighboring models.

\subsection{MTD Module}
An MTD module can be characterized by a triplet $<C, M, T>$, where $C$ signifies the configuration space within which the MTD module is capable of moving, $M$ indicates the strategy employed for the action, and $T$ represents the timing function that determines the intervals at which actions are executed~\cite{sengupta2020survey}. In essence, the design of an MTD module necessitates addressing three fundamental questions: (1) What to move? (2) How to move? and (3) When to move~\cite{zhuang2014towards}?

\textbf{What to Move?}
The threat model analysis delineated in Section~\ref{sec:threatmodel} reveals that the attackers' knowledge and capabilities are restricted to their local area. Thus, the attack surface for benign nodes mainly comes from the model shared by the malicious nodes and the aggregation functions. This paper posits that the configurations of neighboring connections in benign nodes and aggregation functions can be dynamically moved to mitigate the risks posed by poisoning attacks. By modifying the selection of neighboring nodes, benign nodes can effectively diminish the influence exerted by malicious nodes, thereby altering the network topology of the federation. Additionally, by varying the aggregation function, the complexity of executing a successful attack targeting a specific aggregation method can be significantly increased, thereby enhancing the system's resilience against such threats.

\begin{algorithm}[t]
    \caption{Dynamic Topology Strategy}
    \label{alg:topology_mtd}
    \renewcommand{\algorithmicrequire}{\textbf{Require:}}
    \begin{algorithmic}[1]
        \REQUIRE $L$: neighbor nodes, $C$: all nodes in the network, $\kappa_n$: neighbor amount threshold,
                 $nei\_loss$: neighbor model loss on local validation dataset,
                 $nei\_similarity$: similarity between neighbor model and local model,
                 $\kappa_l$: loss threshold, $\kappa_s$: similarity threshold
        
        \FOR{each $c_i \in C$}
            \IF{$c_i \in L$}
                \IF{$nei\_loss[c_i] > \kappa_l$ or $nei\_similarity[c_i] < \kappa_s$}
                    \STATE $L \gets L - c_i$
                \ENDIF
            \ELSE
                \IF{$nei\_loss[c_i] \leq \kappa_l$ and $nei\_similarity[c_i] \geq \kappa_s$}
                    \STATE $L \gets L + c_i$
                \ENDIF
            \ENDIF
        \ENDFOR        
        \IF{mode is proactive and current\_neighbor < $\kappa_n$}
            \STATE $sorted\_nodes \gets$ sort $C$ by $nei\_similarity$ in descending order
            \FOR{each $c_i \in sorted\_nodes$}
                \IF{$c_i \notin L$}
                    \STATE $L \gets L + c_i$
                    \IF{current\_neighbor $\geq \kappa_n$}
                        \STATE break
                    \ENDIF
                \ENDIF
            \ENDFOR
        \ENDIF
        
    \end{algorithmic}
\end{algorithm}

\textbf{How to Move?}
"How to move" refers to the principles and strategies used to alter system configuration elements to increase unpredictability and confuse attackers. This work develops two MTD strategies: the dynamic topology strategy and the dynamic aggregation algorithm strategy.

\subsubsection{Dynamic Topology Strategy}
The topology strategy dynamically alters communication links between nodes, aiming to disrupt predictable patterns that attackers could exploit and prevent adversaries from persistently targeting specific nodes or communication paths. The topology MTD strategy is outlined in Algorithm~\ref{alg:topology_mtd}.

This algorithm starts by calculating how similar neighboring models are to the local model and evaluating the loss on the local validation dataset. These similarity and loss scores are then used in a DBSCAN clustering algorithm to group models together. This helps to identify potential malicious nodes within the network. 

The algorithm employs proactive and reactive approaches to manage the network topology. In both modes, if multiple clusters are detected (indicating potential malicious nodes), reputation thresholds are dynamically set based on cluster boundaries. Nodes then compare their neighbors' reputation scores to these thresholds, disconnecting from those that fall outside the acceptable range and seeking connections with those that meet the criteria. If only one cluster exists (no malicious nodes detected), the proactive mode uses the local model's loss as a threshold, while the reactive mode remains inactive. Both modes aim to minimize exposure to potentially compromised nodes and maintain a secure network environment.

\begin{algorithm}[t]
    \caption{Dynamic Aggregation Algorithm Strategy}
    \label{alg:aggregation_algorithm_mtd}
    \renewcommand{\algorithmicrequire}{\textbf{Require:}}
    \renewcommand{\algorithmicensure}{\textbf{Ensure:}}
    
    \begin{algorithmic}[1]
        \REQUIRE $A$: aggregation algorithm pool, $M$: aggregated models and weights, $N_m$: malicious nodes
        \ENSURE Updated Aggregator
        
        \STATE $A \gets$ ["Krum", "Median", "TrimmedMean"] \COMMENT {Set of aggregation algorithms}
        \STATE $A\_target \gets$ random choice($A$) \COMMENT {Select new aggregation algorithm}
        \STATE $A\_current \gets A\_target$ \COMMENT {Change the aggregator}
        \STATE $A\_current$.set\_nodes\_to\_aggregate($self.trainSet$) 
        \STATE $A\_current$.setRound($self.round$)
        
        \FOR{$s \in M$.keys()}
        
            \IF{$s \notin N_m$}
                \STATE $(m_s, w_s) \gets M[s]$ \COMMENT {Get model and weights}
                \STATE $A\_current$.addModel(s, $m_s, w_s$) \COMMENT {Add model}
            \ENDIF
        \ENDFOR
    \end{algorithmic}
\end{algorithm}

\subsubsection{Dynamic Aggregation Algorithm Strategy}
The dynamic aggregation algorithm strategy enhances the robustness of the DFL system by periodically switching between different aggregation algorithms. Each algorithm has unique strengths and weaknesses against different attacks. By continuously shuffling the strategy, the system remains resilient and reduces the success rate of potential attacks. The algorithm pool includes FedAvg, Krum, Median, and TrimmedMean. The dynamic aggregation algorithm strategy is outlined in Algorithm~\ref{alg:aggregation_algorithm_mtd}.

The MTD strategy operates in two modes. In the proactive mode, benign nodes randomly select an algorithm from the pool in each round. In the reactive mode, if a non-malicious node detects a malicious one, it randomly selects an algorithm and excludes the malicious node from the aggregation process. This dynamic approach helps maintain the system's security and integrity by adapting to potential threats.

\begin{algorithm}[t]
    \caption{Dynamic Reputation Threshold}
    \label{alg:dynamic_reputation_threshold}
    \renewcommand{\algorithmicrequire}{\textbf{Require:}}
    
    \begin{algorithmic}[1]
        \REQUIRE $M'_i$: neighbor models, $m_i$: local model, $sensitive$: sensitivity parameter, $D_i$: local validation dataset
        
        \STATE Initialize $trigger \gets$ False, $similarity\_threshold \gets 0$, $loss\_threshold \gets 10$
        \STATE $nei\_similarity \gets [ \quad ]$
        \STATE $nei\_loss \gets [ \quad ]$
        
        \FOR{each $m_j \in M'_i$}
            \STATE $s_{ij} \gets$ EuclideanDistance($m_i, m_j$)
            \STATE Append $s_{ij}$ to $nei\_similarity$
            \STATE $Loss_{ij} \gets$ reputation\_loss($m_j, D_i$)
            \STATE Append $Loss_{ij}$ to $nei\_loss$
        \ENDFOR
        
        \STATE Apply DBSCAN on $nei\_similarity$ with eps = $sensitive$ to get clusters
        \IF{number of unique clusters $> 1$}
            \STATE $trigger \gets$ True
            \STATE Calculate $similarity\_threshold$ as the mean of the highest lower bound and the lowest upper bound of the clusters
        \ENDIF
        
        \STATE Apply DBSCAN on $nei\_loss$ with eps = $sensitive$ to get clusters
        \IF{number of unique clusters $> 1$}
            \STATE $trigger \gets$ True
            \STATE Calculate $loss\_threshold$ as the mean of the highest lower bound and the lowest upper bound of the clusters
        \ENDIF
        
        \IF{only one unique cluster in both similarity scores and loss scores}
            \STATE $loss\_threshold \gets$ loss score of $m_i$
        \ENDIF
        
        \RETURN $similarity\_threshold, loss\_threshold, trigger$
    \end{algorithmic}
\end{algorithm}

\textbf{When to Move?} 
In the MTD module, "when to move" pertains to selecting the optimal moment to execute the strategies, transitioning from the existing state to an alternative configuration. 
This work proposes two modes of MTD: proactive and reactive. In the proactive mode, the benign node modifies its connected neighbors and the aggregation algorithm in each round. Conversely, in the reactive mode, the benign node requires a triggering event before initiating an appropriate response.

In conventional reactive MTD systems, the trigger of a strategy is primarily contingent upon a pre-established threshold~\cite{feng2023voyager}. However, this pre-established threshold necessitates a prior knowledge, which is often difficult to achieve in practice. Therefore, this paper proposes a dynamic threshold determination algorithm to automate acquiring the threshold for triggering MTD strategies. As shown in Algorithm~\ref{alg:dynamic_reputation_threshold}, this work utilizes the Density-Based Spatial Clustering of Applications with Noise (DBSCAN)~\cite{ester1996density} algorithm to dynamically find the model similarity threshold, loss threshold, and the MTD trigger.

DBSCAN, an unsupervised ML algorithm, clusters data based on its density. It groups densely packed points and identifies outliers in sparse regions, making it valuable for analyzing complex or unknown data distributions. In this work, DBSCAN clusters the reputation scores of neighboring nodes. Despite being trained on non-IID data, benign models typically cluster together due to their inherent consistency. Malicious models, however, deviate significantly, creating distinct clusters. Multiple clusters in model similarity or loss indicate the presence of malicious nodes. By dynamically adjusting reputation thresholds based on these clusters, the system improves its ability to detect and mitigate attacks.

\section{Deployment and Experiments}
\label{sec:exp}
Fedstellar~\cite{beltran2024fedstellar} was selected as the platform for deploying and evaluating the MTD-based approach safeguarding DFL models against poisoning attacks. This platform supports the training of FL models in decentralized and centralized configurations. It facilitates the deployment of the desired scenarios to designated clients and manages the network dynamics among participants. Users can execute DFL scenarios on specific devices or within a virtualized environment utilizing Docker containers.

\subsection{Deployment Configuration}
This work presents the implementation of the MTD-based module outlined in Section~\ref{sec:approach}, utilizing the Python programming language. The experimental configurations are detailed in \tablename~\ref{tab:exp_config}. This experiment aims to assess the performance of the proposed MTD module by conducting a comparative analysis with various baseline models across multiple datasets and data distributions. Additionally, different MTD modes and strategies are evaluated to examine their respective strengths and weaknesses.

\begin{table}[t]
\centering
\caption{Experiments Configuration}
\label{tab:exp_config}
\resizebox{\columnwidth}{!}{%
\begin{tabular}{l|l}
\toprule
\textbf{CONFIGURATION} & \textbf{VALUE} \\ \midrule
Dataset & MNIST, FashionMNIST, CIFAR10 \\ \hline
Dirichlet $\alpha$ & 0.5 \\ \hline
Attack & Label Flipping, Model Poisoning, Backdoor \\ \hline
PNR & 10\%, 30\%, 50\%, 70\% \\ \hline
Topology & Fully-Connected, Ring, Star \\ \hline
Baseline Algorithm & FedAvg, Krum, Median, TrimmedMean \\ \hline
 & Reactive Topology MTD Only, \\ 
 MTD Strategy& Reactive Aggregation MTD Only, \\ 
 & Reactive Topology \& Aggregation MTD, \\ 
 & Proactive Topology \& Aggregation MTD \\ \midrule
\end{tabular} 
}
\end{table}

\textbf{Datasets and Models.}
Three widely utilized benchmark datasets~\cite{feng2024dart}, along with their corresponding models, have been selected to assess the performance of the proposed MTD-based module:
\begin{itemize}
    \item \textbf{MNIST}~\cite{mnistwebsite} is a widely-used handwritten digits (0-9) dataset for image recognition, consisting of 60,000 training images and 10,000 testing images, each in gray-scale and sized 28x28 pixels. A Multilayer Perceptron (MLP) model is used, consisting of two fully connected hidden layers with 256x128. The MLP is trained for 3 epochs per federated round, optimized using Adam with a learning rate of \(1 \mathrm{e}{-3}\), and cross-entropy loss is applied.
    \item \textbf{FashionMNIST}~\cite{xiao2017fashion} is a dataset created as a more challenging alternative to MNIST for benchmarks. It contains 60,000 training and 10,000 testing 28x28 gray-scale images across 10 fashion categories (e.g., trousers and coats). A small CNN with two convolutional layers (32 and 64 filters) is employed, followed by two fully connected layers. The model is trained on FashionMNIST for 3 epochs per round using Adam optimizer with a learning rate of \(1 \mathrm{e}{-3}\).
    \item \textbf{CIFAR10}~\cite{krizhevsky2009learning} is a benchmark dataset for ML, consisting of 50,000 training and 10,000 testing images (32x32 pixels) across 10 classes (e.g., airplane, automobile, bird). The colorful images and diverse classes make it more challenging than MNIST and FashionMNIST. The SimpleMobileNet \cite{sinha2019thin} model is utilized. It starts with a convolutional layer (32 filters, 3x3 kernel), followed by max pooling. Then, five depthwise separable convolutional layers are used, followed by adaptive average pooling and a dense layer. The model is trained using Adam optimizer (learning rate \(1 \mathrm{e}{-3}\)) for 5 epochs per round.
\end{itemize}

Experiments have been conducted in non-IID settings. The datasets were distributed according to the Dirichlet distribution with $\alpha$ parameters set to 0.5.

\textbf{Attack Setups.}
Three different poisoning attacks have been designed and implemented~\cite{feng2024dart}, respectively:
\begin{itemize}
    \item \textbf{Untargeted Label Flipping.} In this attack, malicious participants alter all of the training data samples' labels to incorrect values.
    \item \textbf{Model Poisoning.} In this attack, malicious participants inject 50\% of the Gaussian noise into their local model updates and share these corrupted model parameters with their neighbors.
    \item \textbf{Backdoor.} The trigger developed in this experiment involved incorporating a 10x10 watermark in the upper left corner of the sample images corresponding with the fifth class across each dataset. During the testing phase, any images displaying a watermark in the upper left corner will be classified as belonging to the fifth class.
\end{itemize}
This experiment further examined the influence of varying threat levels on the performance of the baseline algorithm and the MTD module by manipulating the proportions of compromised nodes, denoted as the Poisoned Node Ratio (PNR). Specifically, PNR values of 10\%, 30\%, 50\%, and 70\% were employed to characterize environments with low, medium, and high levels of compromise.

\textbf{Federation Setups.}
The experiments were performed using a federation with 10 nodes with 10 rounds. The initial network topology varied among fully connected ring and star structures.
The network topology could change dynamically during training if the topology MTD was enabled.


During the DFL learning process, each node acted as a trainer and an aggregator. Different baseline aggregation algorithms, including FedAvg, Krum, Median, and TrimmedMean, were compared with the proposed MTD-based approach. Four different combinations of MTD modes and strategies were tested, including reactive dynamic aggregation strategy, reactive dynamic topology + aggregation strategy, proactive dynamic topology + aggregation strategy, and proactive dynamic topology + aggregation strategy.

\subsection{Experiments}
This experiment first evaluates the performance of different algorithms across multiple topologies when there is no attack to compare their robustness against poisoning attacks comprehensively.

\begin{table}[b]
\setlength{\tabcolsep}{3pt}
\caption{F1-Score for MNIST, FashionMNIST, and CIFAR10  without Attacks in Fully, Ring and Star Topology}
\label{tab:baseline_performance}
\resizebox{\columnwidth}{!}{%
\begin{tabular}{c|p{1.1cm}|lll|lll|lll}
\toprule
\multicolumn{1}{l}{\textit{\textbf{}}} & \textbf{} & \multicolumn{3}{c|}{\textbf{MNIST}} & \multicolumn{3}{c|}{\textbf{FashionMNIST}} & \multicolumn{3}{c}{\textbf{CIFAR10}} \\
\midrule
\textit{Non-IID $\alpha$} &   \textit{Agg.}   & Fully & Ring  & Star  & Fully & Ring  & Star  & Fully & Ring  & Star  \\ \hline
\multirow{4}{*}{100} & FedAvg      & 0.963 & 0.952 & 0.958 & 0.893 & 0.876 & 0.877 & 0.729 & 0.660 & 0.695 \\
                     & Krum        & 0.955 & 0.951 & 0.956 & 0.879 & 0.871 & 0.881 & 0.693 & 0.642 & 0.686 \\
                     & Median      & 0.960 & 0.949 & 0.958 & 0.892 & 0.853 & 0.875 & 0.730 & 0.629 & 0.707 \\
                     & Trimmed-Mean & 0.965 & 0.939 & 0.957 & 0.895 & 0.858 & 0.881 & 0.728 & 0.629 & 0.691 \\ \hline
\multirow{4}{*}{0.5} & FedAvg      & 0.919 & 0.867 & 0.894 & 0.818 & 0.781 & 0.792 & 0.584 & 0.445 & 0.487 \\
                     & Krum        & 0.847 & 0.852 & 0.872 & 0.759 & 0.752 & 0.785 & 0.457 & 0.436 & 0.458 \\
                     & Median      & 0.910 & 0.848 & 0.878 & 0.803 & 0.751 & 0.805 & 0.552 & 0.434 & 0.487 \\
                     & Trimmed-Mean & 0.919 & 0.843 & 0.886 & 0.830 & 0.744 & 0.788 & 0.553 & 0.435 & 0.475 \\ \midrule
\end{tabular}%
}
\end{table}

As illustrated in \tablename~\ref{tab:baseline_performance}, factors such as the aggregation algorithms, data distribution, and network topology significantly influence the performance of the DFL model. In general, the FedAvg and TrimmedMean algorithms demonstrate superior performance in a non-attack scenario, followed by the Median and Krum algorithms. The models exhibit enhanced performance within a fully connected topology, while a decline in effectiveness is observed under Star and Ring topologies. This suggests that a denser network topology facilitates more efficient information exchange among DFL nodes. Moreover, the model's performance deteriorates as the degree of non-IID data increases (i.e., alpha value decreases). With no attacks, and when alpha is equal to 0.5, the average F1-Score of the model is about 0.91 on the MNIST dataset, 0.8 on FashionMNIST, and 0.55 on CIFAR10. On the MNIST and FashionMNIST datasets, when alpha decreases from 100 to 0.5, the F1-Score decreases by about 0.05. However, on the CIFAR10 dataset, there is a decrease of about 0.15, indicating that non-IID significantly impacts the effectiveness of the DFL model. 

This result establishes a foundational benchmark in the absence of an attack. Additionally, it shows that the robustness of the DFL model can be influenced by moving the topology and aggregation algorithms. Furthermore, it underscores the necessity for optimizing defenses in non-IID scenarios. Subsequent experiments are conducted within non-IID scenarios, with an alpha value of 0.5.
\begin{figure}[t]
    \centering
    \includegraphics[width=0.48\textwidth]{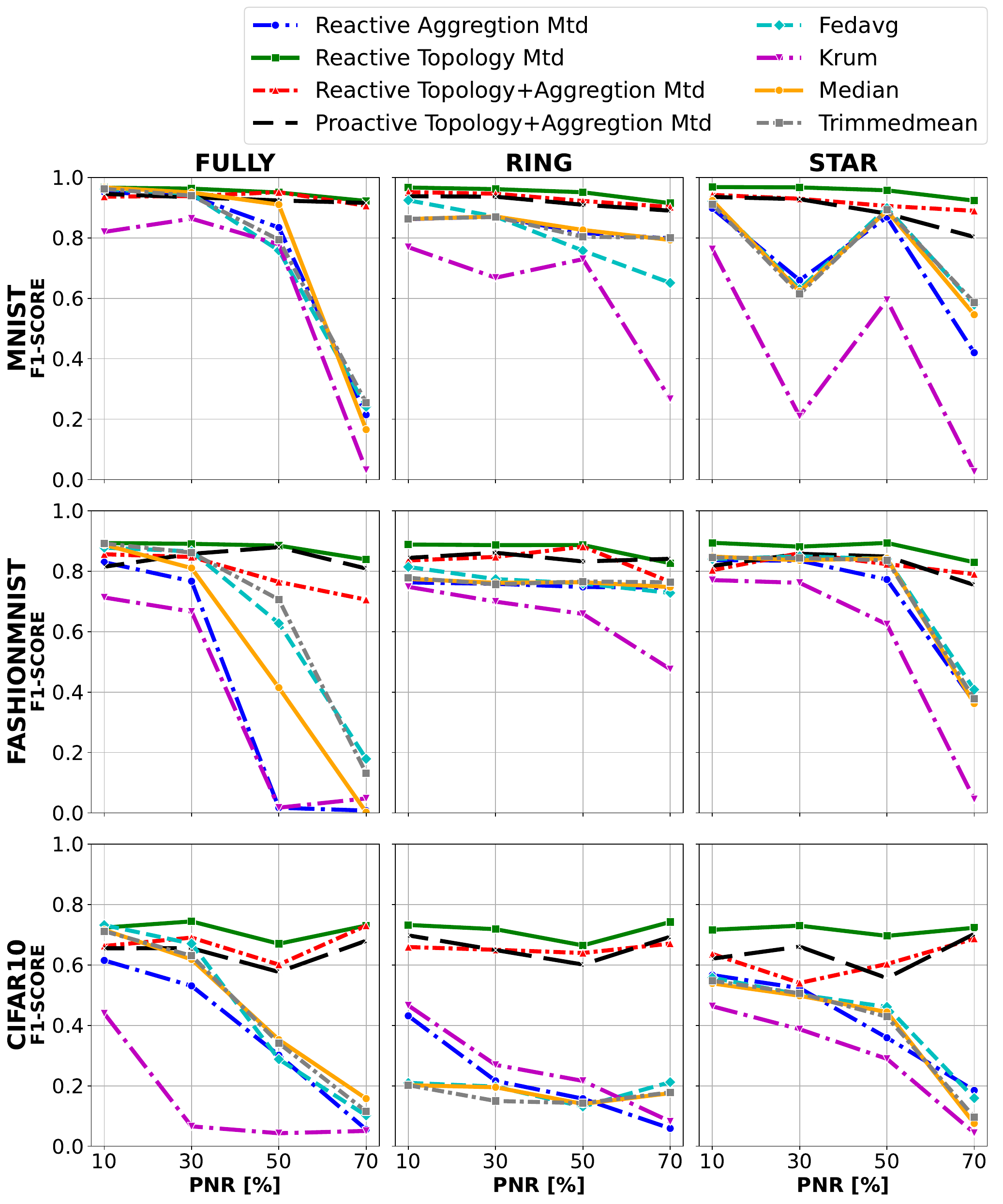}
    \caption{Average F1-Score Results for Label Flipping}
    \label{fig:labflipping}
\end{figure}

\textbf{Untargeted Label Flipping.}
In this experiment, adversarial participants manipulate the local training dataset by arbitrarily changing the labels associated with their data samples. This study assesses defense mechanisms against untargeted label flipping using the average F1-Score of benign nodes during the final iteration of the DFL process as the evaluation metric. An elevated F1-Score suggests that the benign nodes experience minimal disruption from malicious nodes, thereby indicating greater effectiveness of the defense mechanisms.

The experiments compare the defense effectiveness of the proposed MTD strategies and the baseline algorithms in three datasets with three different topologies, and the results are shown in \figurename \ref{fig:labflipping}. In low-threat environments (PNR equal to 10\%), the MTD strategies and the baseline algorithms effectively mitigate label-flipping attacks. However, as the PNR increases to 50\%, the baseline algorithms' performance declines across all three datasets with three topologies. When the PNR reaches 70\%, the average F1 Scores for the FedAvg, Krum, Median, and TrimmedMean algorithms plummet to below 0.2 for all datasets within a fully connected topology. This significant reduction indicates that malicious nodes have substantially compromised the integrity of the benign nodes.

Conversely, MTD-based strategies, particularly those employing dynamic topologies, whether reactive or preventive, prove effective in safeguarding benign nodes. Even at a PNR of 70\%, enforcing dynamic topology strategies yields average F1-Scores exceeding 0.9 for the MNIST dataset, 0.8 for FashionMNIST, and 0.7 for CIFAR10. However,  strategies relying solely on dynamic aggregation algorithms do not exhibit a comparative advantage, delivering results similar to baseline algorithms. In conclusion, dynamic topology-based MTD strategies, encompassing both reactive and preventive modes,  can effectively mitigate label flipping attacks.

\begin{figure}[t]
    \centering
    \includegraphics[width=0.48\textwidth]{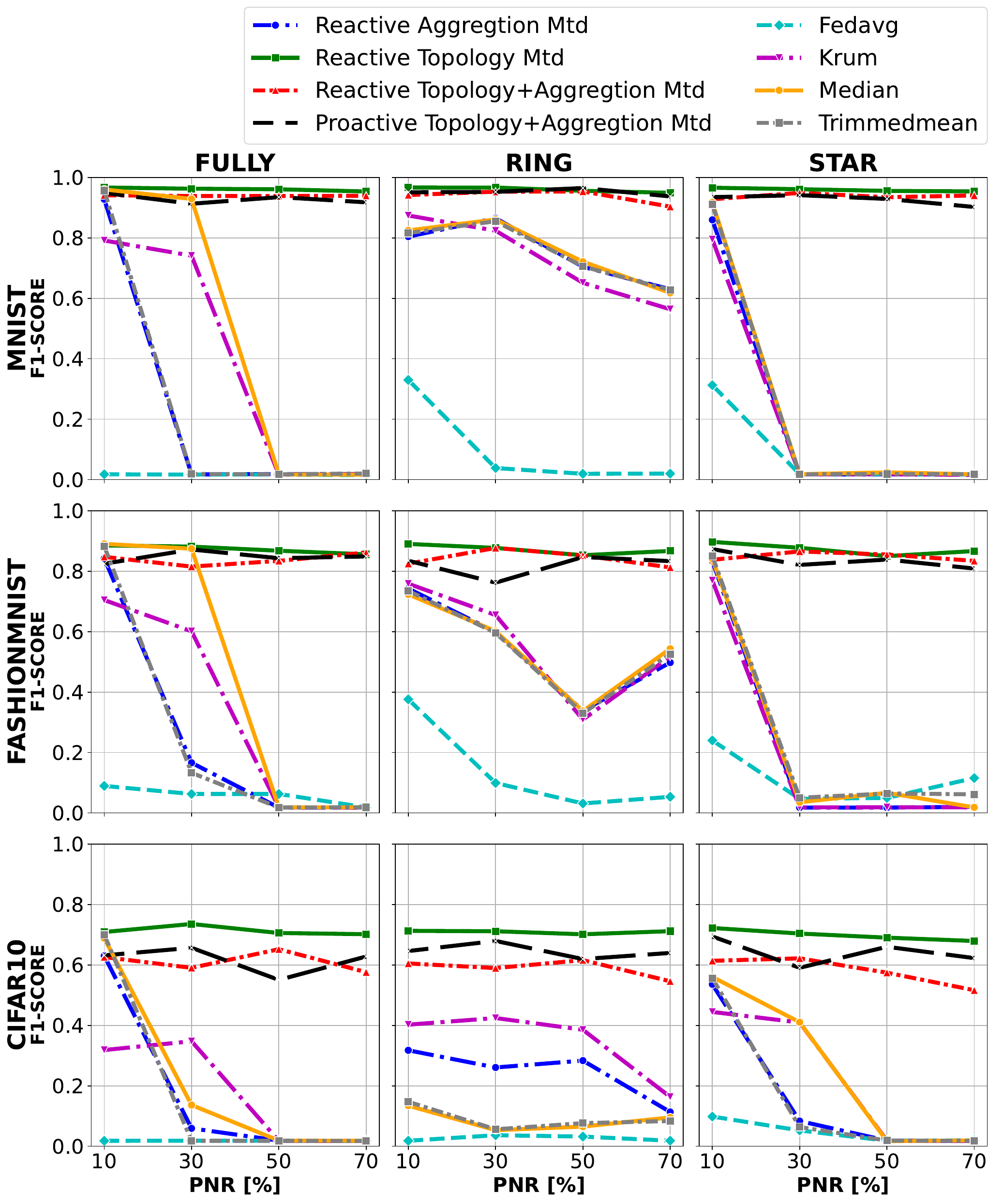}
    \caption{Average F1-Score Results for Model Poisoning}
    \label{fig:modelpoisoning}
\end{figure}

\textbf{Model Poisoning.} 
In this attack, the model weights in malicious nodes are randomly changed by Gaussian noise before sending to the neighbors in each round. As with label flipping attacks, the average F1-Score is used to evaluate the effectiveness of the defense algorithms, and the results are shown in \figurename \ref{fig:modelpoisoning}.

Similar to label flipping attacks, baseline defenses are effective when the PNR equals 10\%. However, it is noteworthy that the FedAvg algorithm shows significant vulnerability even at a PNR of 10\%. When the PNR increases to 30\%, the performance of the baseline algorithms drops dramatically. When the PNR is increased to 50\%, the baseline algorithms' performance drops to nearly 0.1 on all three datasets, indicating that these defenses cannot mitigate poisoning attacks in such scenarios. In contrast, the defense strategies using MTD, especially the dynamic topology defense, can protect the DFL from model poisoning attacks. The F1-Score stays at 0.95 on the MNIST, 0.82 on the FashionMNIST, and 0.7 on the CIFAR10, even though the PNR grows from 10\% to 70\%.

Given that the attack strategy employed in this experiment is a random weight attack, the dynamic aggregation algorithm fails to demonstrate its advantage. To summarize, the defense strategies based on MTD effectively mitigate model poisoning attacks within non-IID environments.

\begin{figure}[t]
    \centering
    \includegraphics[width=0.48\textwidth]{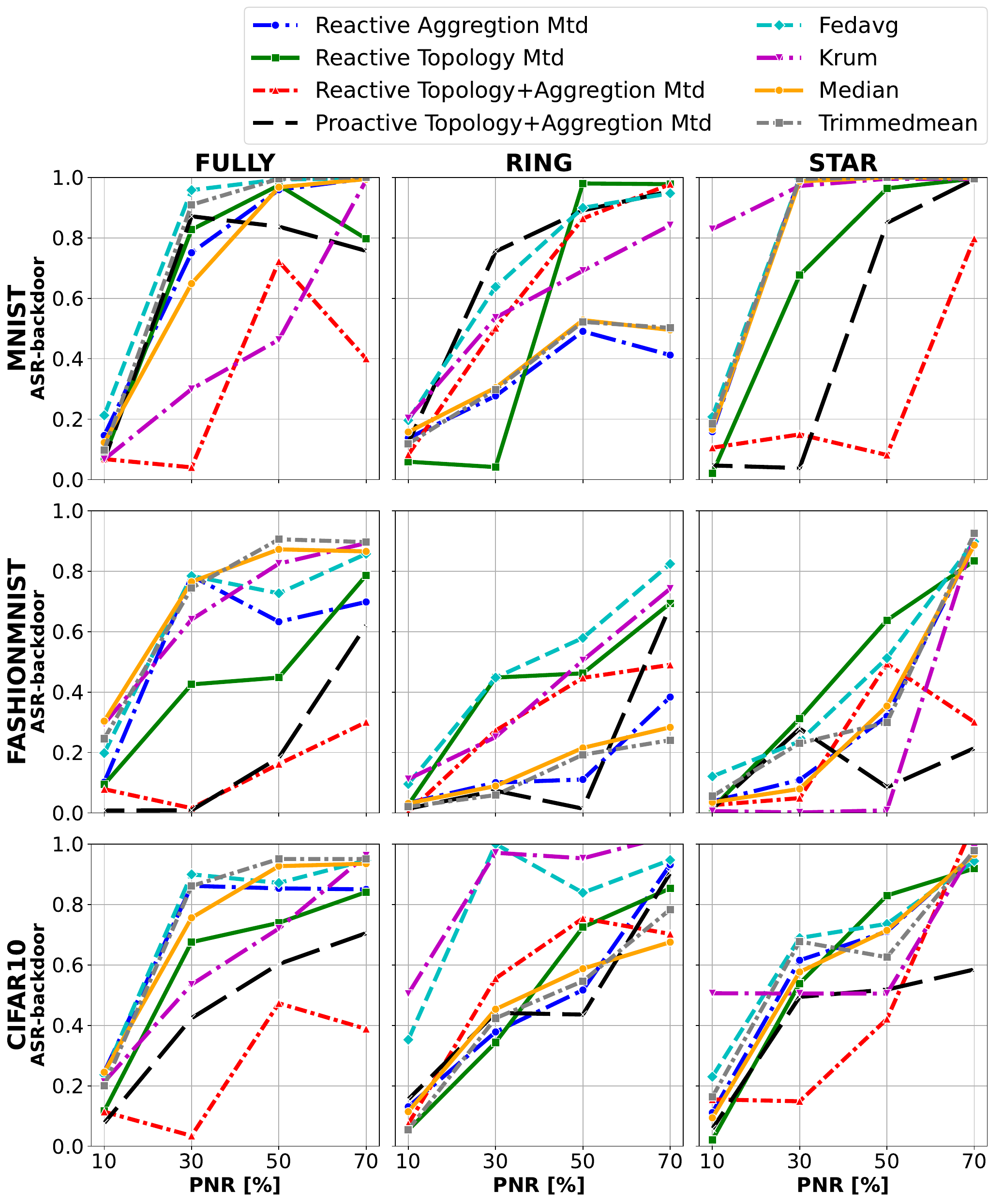}
    \caption{Average ASR-backdoor Results for Backdoor}
    \label{fig:backdoor}
\end{figure}

\textbf{Backdoor.}
In backdoor attacks, adversarial participants endeavor to induce the misclassification of inputs to a specified target label \( l_{t} \) through the deployment of a synthetic trigger. In this experiment, this trigger is exemplified by the depiction of an "X" on images drawn from three datasets. The performance of defense mechanisms is assessed using the Attack Success Rate (ASR) metric~\cite{feng2024sentinel}, as delineated in Equation \eqref{eq:BA}. Here, \( c_{ij} \) denotes the count of instances with a true label \( y_i \) that are predicted to belong to label \( \hat{y}_j \). The variable \( L \) signifies the collection of labels present within the respective dataset, while \( B \) denotes the subset of data specifically associated with the backdoor attack.

\begin{equation} \label{eq:BA}
    ASR = \frac{\sum_{j=0}^{|L|} c_{j, t} - c_{t, t}}{|B| - c_{t, t}}
\end{equation}

A higher ASR means that more malicious samples are misclassified, i.e., the backdoor is successfully implanted, thus indicating a poorer defense. A good defense against backdoor attacks should be able to reduce the ASR. \figurename \ref{fig:backdoor} presents the results of evaluating backdoor attacks under three topologies on three datasets.

As illustrated in \figurename \ref{fig:backdoor}, the ASR increases with the increase of PNR for the baseline algorithms and the MTD-based algorithms, indicating the limitations of the MTD algorithms proposed in this paper when mitigating the backdoor attacks. Notably, although these MTD algorithms do not entirely succeed in countering the backdoor attack, the defense strategy that integrates a dynamic aggregation function and dynamic topology in both proactive and reactive modes demonstrates the lowest ASR across the three datasets utilizing fully connected and star topologies. This result suggests that the MDT-based approach has advanced over the baseline algorithms in combating backdoor attacks.

\begin{figure}[t]
    \centering
    \includegraphics[width=0.48\textwidth]{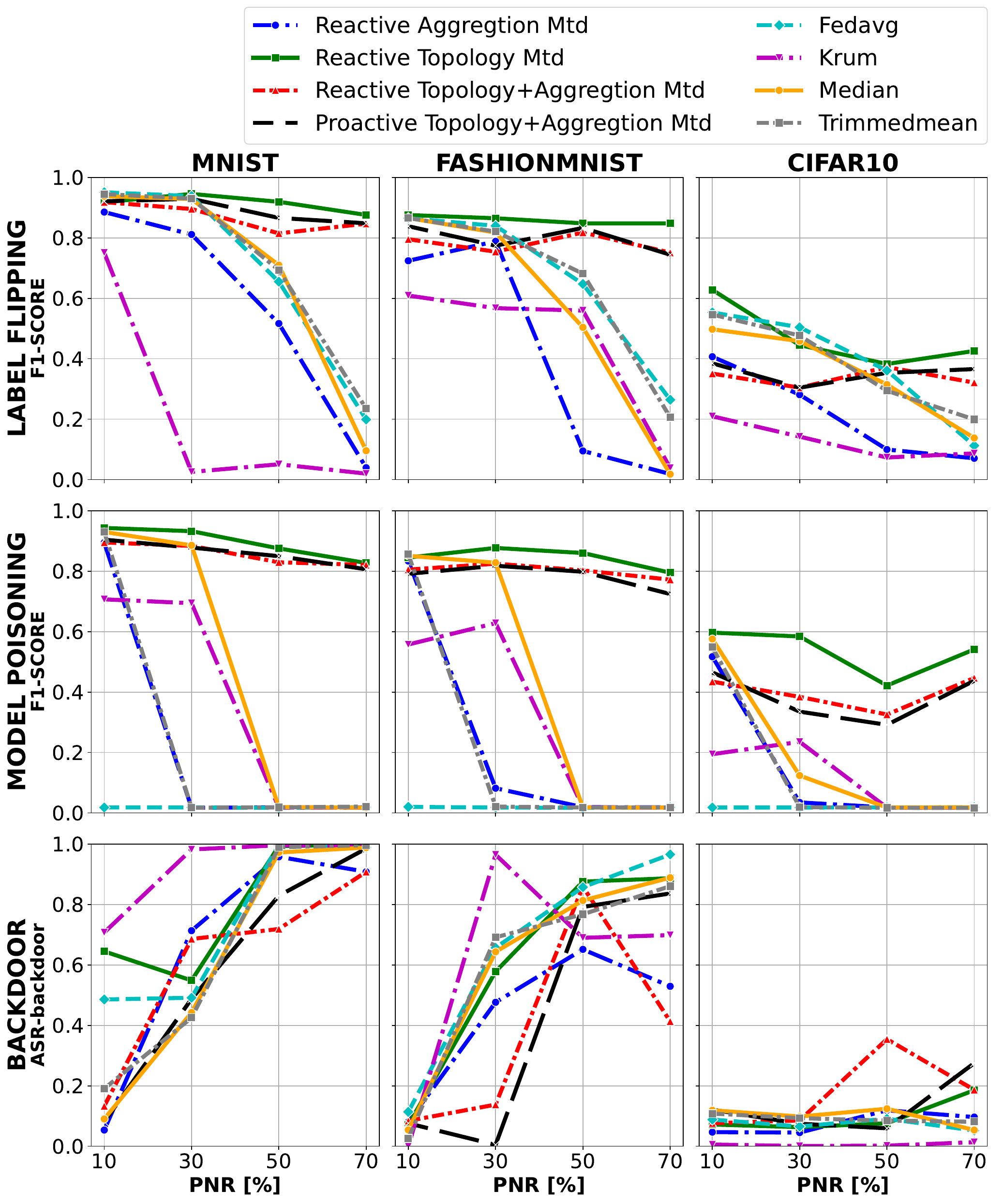}
    \caption{Performance of the Algorithms against poisoning attack on three Datasets with Fully connected topology when Dirichlet $\alpha$=0.1}
    \label{fig:non_iid_01}
\end{figure}

\textbf{Extreme Non-IID Scenario.}
To investigate the effectiveness of the algorithms proposed in this study within extreme non-IID environments, this experiment is established to evaluate the performance of the reference algorithms alongside the MTD algorithms. This assessment is conducted under various attack scenarios within a fully-connected network, utilizing a Dirichlet $\alpha$ parameter of 0.1, as depicted in \figurename \ref{fig:non_iid_01}.

The results indicate a general decline in the performance of all models across the three datasets, attributable to the heightened levels of non-IID conditions. Furthermore, regarding label flipping and model poisoning attacks, the results reveal that the extent of non-IID does not significantly impact the effectiveness of the MTD algorithms. Specifically, the defenses employing the reactive dynamic topology strategy and those integrating both dynamic topology and dynamic aggregation functions outperform the other algorithms.

\begin{table}[t]
\setlength{\tabcolsep}{3pt}
\caption{F1 Score with Reactive Dynamic Topology MTD across Three Datasets, with a PNR Rate of 30\%, Alpha Value of 0.5, and Fully Connected Topology}
\label{tab:20nodes}
\resizebox{\columnwidth}{!}{%
\begin{tabular}{l|l|lll}
\toprule
\textit{Non-IID $\alpha$} &   \textit{Attack}   & \textit{MNIST} & \textit{FashionMNIST}  & \textit{CIFAR10} \\ \hline
0.5 & Lable Flipping      & 0.896$\pm$0.024 & 0.784$\pm$0.099 & 0.494$\pm$0.042 \\
      & Model Poisoning     & 0.879$\pm$0.021 & 0.799$\pm$ 0.039 & 0.511$\pm$0.069 \\\midrule
\end{tabular}%
}
\end{table}

\textbf{Scalability.} To investigate the scalability of the algorithm presented in this paper, an experimental setup is conducted utilizing 20 nodes with a fully connected topology. This scenario involved simulating an environment where 30\%  of the nodes are attacked. The performance of the reactive dynamic topology algorithm is evaluated under two specific attack types: label flipping and model poisoning. A Dirichlet $\alpha$ parameter of 0.5 is utilized for this assessment. The results are summarized in \tablename~\ref{tab:20nodes}, indicating that the federated approach utilizing the MTD algorithm remains resilient against poisoning attacks in a large network. This observation underscores the adaptability and scalability of the proposed algorithms across networks of varying sizes.

\section{Discussion}
\label{sec:discussion}
Extensive experiments have demonstrated that the proposed MTD-based approach can effectively mitigate poisoning attacks in non-IID environments. 

\textbf{\textit{A. Reduced effectiveness in the backdoor.}} Nevertheless, despite this approach demonstrating superiority over existing state-of-the-art algorithms, there is potential for enhancement in addressing targeted attacks, such as backdoor attacks. To investigate the reasons behind the sub-optimal performance of the MTD-based approach in backdoor attacks, this work analyzes the distribution of reputation scores generated by benign nodes. It is conducted under conditions where the PNR is set at 10\% and within a fully connected topology. The comparison is performed across three datasets, focusing on model poisoning and backdoor attack scenarios.

\begin{figure}[b]
    \centering
    \includegraphics[width=0.8\columnwidth]{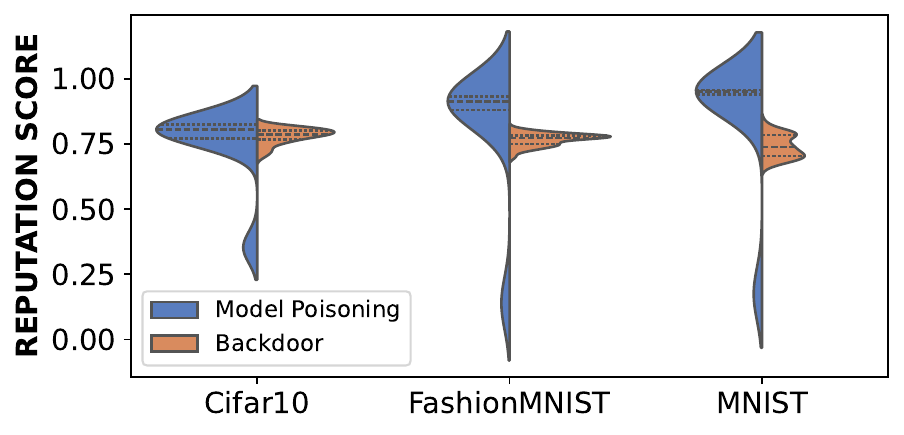}
    \caption{Distribution of Reputation Score on MNIST, FashionMNIST, CIFAR10 dataset when PNR equals 10\%}
    \label{fig:reputation}
\end{figure}

As illustrated in \figurename \ref{fig:reputation}, the reputation score assigned to benign nodes about their neighboring nodes exhibits a noticeable polarization in the model poisoning attack scenario. Specifically, the reputation score for benign nodes is significantly higher than for malicious nodes. In contrast, this polarization is absent in backdoor attacks, where the distribution of the reputation score is more centralized. Notably, the reputation score enforces the MTD strategies. Thus, the effectiveness of MTD strategies is diminished in backdoor attack scenarios.

\textbf{\textit{B. Enhanced performance in non-IID settings.}} It is essential to highlight that the proposed solution enhances the model performance of DFL in non-IID environments, especially when using the dynamic topology strategy. The average F1-Score achieved by the proposed approach reaches 0.95 for the MNIST dataset, 0.83 for FashionMNIST, and 0.70 for CIFAR10 when using the reactive dynamic topology strategy, as illustrated in \figurename \ref{fig:labflipping} and \ref{fig:modelpoisoning}. These results indicate that the algorithm not only mitigates the impacts of poisoning attacks but also enhances model performance in non-IID settings. The effectiveness of the dynamic topology lies in its tendency to facilitate connections and aggregations among nodes that exhibit high similarity, diminishing the impact of non-IID data.

\textbf{\textit{C. Trade-off.}} Regarding trade-off, the algorithm proposed in this paper increases the computational complexity compared to reference models, but the increased complexity is within polynomial complexity. However, the reference algorithms, including Krum, also need to calculate the distance of the models. Therefore, when evaluating the advantages in terms of security afforded by implementing MTD, this study posits that the additional computational complexity is justifiable. Regarding network overhead, the algorithms introduced herein do not necessitate any supplementary transmission of information, thus avoiding any additional burden on network overhead.

\section{Conclusion and Future Work}
\label{sec:conclusion}
This work proposes an MTD-based approach to mitigate poisoning attacks in DFL, especially in non-IID environments. It introduces a dual-reputation score mechanism assessing model similarity and loss, with proactive and reactive MTD modes using dynamic topology and aggregation strategies. By leveraging the DBSCAN clustering algorithm, the approach dynamically adjusts thresholds, overcoming the limitations of fixed threshold defenses. The dynamic topology enhances node connections aligned with local models. Experiments on MNIST, FashionMNIST, and CIFAR10 datasets demonstrate the approach's effectiveness against untargeted attacks and superior performance over state-of-the-art aggregation algorithms, offering a robust security framework for non-IID data.

Future research will investigate integrating additional MTD strategies to comprehensively address the diverse landscape of poisoning attacks and enhance resilience against optimized and dynamic attack methodologies. Further optimization of the MTD-based approach, particularly in the context of backdoor attacks, will also be explored. Additionally, the research scope will be extended to related domains, including the application of the MTD paradigm to counter various other types of attacks targeting DFL, such as Membership Inference Attacks.

\section*{Acknowledgment}
This work was supported partially by (a) the University of Zürich UZH, (b) the Swiss Federal Office for Defense Procurement (armasuisse) with the CyberMind project (CYD-C-2020003).

\bibliographystyle{IEEEtran}
\balance
\bibliography{references}

\end{document}